\documentclass[12pt]{article}

\usepackage{setspace, graphicx, caption, subcaption, float, relsize, rotating, color, verbatim, multirow}
\usepackage[round, comma, longnamesfirst]{natbib}

\usepackage{graphicx}   
\usepackage[colorlinks=true, urlcolor=blue,linkcolor=blue, citecolor=magenta]{hyperref}

 \renewcommand{\thefootnote}{\fnsymbol{footnote}}

\usepackage{empheq}
\usepackage{enumerate} 

\usepackage{dsfont}
\usepackage{mathtools}
\usepackage{bm}

\usepackage{stmaryrd}
\usepackage{amssymb} 
\usepackage{amsmath}
\usepackage{amsfonts}
\usepackage{amsthm} 
\usepackage{mathrsfs}
\usepackage{bigints} 
\usepackage{enumerate} 
\usepackage{authblk} 
\usepackage{color} 
\usepackage{sectsty}

\usepackage{amsfonts}

\usepackage{times}
\usepackage{bm}
\usepackage{natbib}

\usepackage[plain,noend]{algorithm2e}

\newtheorem{thm}{Theorem}[section]

\newtheorem{prop}[thm]{Proposition}

\newtheorem{theorem}[thm]{Theorem}
\newtheorem{assumption}[thm]{Assumption}
\newtheorem{lemma}[thm]{Lemma}

\newcommand{\nc}{\newcommand}

\nc{\dps}{\displaystyle}
\nc{\tr}{\text{tr}}

\title{\Large   A Simple and Efficient Estimation of the Average Treatment Effect in
	the Presence of Unmeasured Confounders } 
\author[$^\star$]{Chunrong Ai \thanks{E-mail:  chunrong.ai@warrington.ufl.edu}}
\author[$^\dagger$]{Lukang Huang \thanks{E-mail:  huanglukang@ruc.edu.cn}}
\author[$^\dagger$]{Zheng Zhang \thanks{E-mail: zhengzhang@ruc.edu.cn}}
\affil[$\star$]{Department of Economics, University of Florida}
\affil[$\dagger$]{Institute of Statistics \& Big Data, Renmin University of China}

\numberwithin{equation}{section}
 
\allowdisplaybreaks 

\begin{document}
 
\setcounter{section}{0}

\maketitle

\baselineskip0.6cm

\begin{abstract}
\cite{wang2016bounded} studied identification and estimation of the average
treatment effect when some confounders are unmeasured. Under their
identification condition, they showed that the semiparametric efficient
influence function depends on five unknown functionals. They proposed to
parameterize all functionals and estimate the average treatment effect from
the efficient influence function by replacing the unknown functionals with
estimated functionals. They established that their estimator is consistent
when certain functionals are correctly specified and attains the
semiparametric efficiency bound when all functionals are correctly
specified. In applications, it is likely that those functionals could all be
misspecified. Consequently their estimator could be inconsistent or
consistent but not efficient. This paper presents an alternative estimator
that does not require parameterization of any of the functionals. We
establish that the proposed estimator is always consistent and always
attains the semiparametric efficiency bound. A simple and intuitive
estimator of the asymptotic variance is presented, and a small scale
simulation study reveals that the proposed estimation outperforms the
existing alternatives in finite samples. 
\end{abstract}

\renewcommand\thefootnote{\arabic{footnote}}

\section{Introduction}
 A common approach to account for individual heterogeneity in the treatment
 effect literature on observational data is to assume that there exist
 confounders, and conditional on these confounders, there is no systematic
 selection into the treatment (i.e., the so-called \emph{Unconfounded
 	Treatment Assignment} condition suggested in \cite{rosenbaum1983central,
 	rosenbaum1984reducing}). Under this assumption, several procedures for
 estimating the average treament effect (hereafter ATE) have been proposed,
 including the weighting procedure (\cite{rosenbaum1987model}, \cite
 {hirano2003efficient}, \cite{tan2010bounded}, \cite{imai2014covariate}, \cite
 {chan2016globally}, \cite{yiu2018covariate}); the matching procedure (\cite
 {rosenbaum2002observational}, \cite{rosenbaum2002covariance}, \cite
 {dehejia1999causal}); and the regression procedure (\cite
 {heckman1997matching}, \cite{heckman1998matching}, \cite{imbens2006mean}, 
 \cite{chen2008semiparametric}). For example survey, see \cite
 {imbens2009recent} and \cite{imbens2015causal}. A critical requirement in
 this literature is that all confounders are observed and available to
 researchers. In applications, however, it is often the case that some
 confounders are either not observed or not availale. In this case, the
 average treatment effect is only partially identified even with the aid of
 some insrumental variables (see \cite{Imbens1994Identification}, \cite
 {Joshua1996Identification}, \cite
 {Abadie2003Semiparametric,Abadie2002Instrumental,Tan2006Regression,Cheng2009Efficient,Ogburn2015Doubly}
 for examples).\newline
 
 Recently \cite{wang2016bounded} suggested a noval identification condition
 of ATE when some confounders are not available. Under their condition, they
 showed that the semiparametric efficient influence function of ATE depends
 on five unknown functionals. They proposed to parameterize all five
 functionals, estimate those functionals with appropriate parametric
 approaches, plug the estimated functionals into the influence function, and
 then estimate the ATE from the estimated influence function. They
 established that their estimator is consistent if certain functionals are
 correctly parameterized and attains the semiparametric efficiency bound if
 all functionals are correctly specified. In applications, it is quite
 possible\ that some or all of the five functionals are misspecified and
 consequently their estimator could be inefficient or worse, inconsistent.
 This paper proposes an alternative, intuitive and easy to compute estimation
 that does not require parameterization of any of the five unknown
 functionals. We estbalish that under some sufficient conditions the proposed
 estimator is consistent, asymptotically normally distributed and attains the
 semiparametric efficiency bound. Moreover, the proposed procedure provides a
 natural and convenient estimate of the asymptotic variance. \newline
 
 The paper is organized as follows. Section \ref{sec:framework} describes the
 basic framework. Section \ref{sec:estimation} describes the proposed
 estimation and derives the large sample properties of the proposed
 estimator. Section \ref{sec:variance} presents a consistent variance
 estimator. Since the proposed procedure depends on smoothing parameters,
 Section \ref{sec:K} presents a data driven method for selecting the
 smoothing paprameters. Section \ref{sec:simulation} reports a small scale
 simulation study to evaluate the finite sample performance of the proposed
 estimator. 
Some concluding remarks are in Section \ref%
 {sec:discussion}. All technical proofs are relegated to the Appendix and the
 supplementary material.

\section{ Basic Framework}\label{sec:framework}
Let $D\in \{0,1\}$ denote the binary treatment
indicator, and let $Y(1)$ and $Y(0)$ denote the potential outcomes when an
individual is assigned to the treatment and control group respectively. The
parameter of interest is the population average treatment effect $\tau =%
\mathbb{E}[Y(1)-Y(0)]$. Estimation of $\tau $ is complicated by the presence
of confounders and the fact that $Y(1)$ and $Y(0)$ cannot be observed
simultaneously. To distinguish observed confounders from unobserved
confounders, we shall use $X$ to denote the observed confounders and use $U$
to denote the unmeasured confounders. It is well established in the
literature that, when all confounders are observed, the following \emph{%
	Unconfounded Treatment Assignment} condition is sufficient to identify $\tau $%
:

\begin{assumption}  \label{as:CIA}
	$(Y(0),Y(1))\perp (D,Z)|(X,U)$.
\end{assumption}

When $U$ is unmeasured, we have the classical omitted variable problem,
causing the treament indicator $D$ to be endogenous. To tackle the
endogeneity problem, instrumental variable is often the preferred choice.
Let $Z\in \{0,1\}$ denote the variable satisfying the following classical
instrumental variable conditions:

\begin{assumption}[Exclusion restriction] \label{as:exclusion}	$\forall z,d$, $Y(z,d)=Y(d)$, where $Y(z,d)$ is the response that would be observed if a unit were exposed to $d$ and
	the instrument had taken value $z$ to be well defined.
\end{assumption}

\begin{assumption}[Independence] \label{as:independence}
	$Z\perp U|X$.
\end{assumption}

\begin{assumption} [IV relavance] \label{as:relavance}
	$Z \not\perp D|X$.
\end{assumption}

\cite{wang2016bounded} showed that Asssumptions \ref{as:CIA}- \ref%
{as:relavance} alone do not identify $\tau $, but if in addition one of the
following conditions holds: \label{as:U-ZD}

\begin{enumerate}
	\item there is no additive $U$-$Z$ interaction in $\mathbb{E}[D|Z,X,U]$: 
	\begin{equation*}
	\mathbb{E}[D|Z=1,X,U]-\mathbb{E}[D|Z=0,X,U]=\mathbb{E}[D|Z=1,X]-\mathbb{E}
	[D|Z=0,X]\ .
	\end{equation*}
	
	\item there is no additive $U$-$d$ interaction in $\mathbb{E}[Y(d)|X,U]$: 
	\begin{equation*}
	\mathbb{E}[Y(1)-Y(0)|X,U]=\mathbb{E}[Y(1)-Y(0)|X]\ ,
	\end{equation*}
\end{enumerate}
then ATE is identified and can be expressed as 
\begin{align}
\label{id:tau}
\tau =\mathbb{E}[\delta (X)]=\mathbb{E}\left[ \frac{\delta ^{Y}(X)}{\delta
	^{D}(X)}\right] \ ,
\end{align}%
where 
\begin{align*}
& \delta ^{Y}(X)=\mathbb{E}[Y|Z=1,X]-\mathbb{E}[Y|Z=0,X]\ , \\
& \delta ^{D}(X)=\mathbb{E}[D|Z=1,X]-\mathbb{E}[D|Z=0,X]\ , \\
& \delta (X)=\delta ^{Y}(X)/\delta ^{D}(X)\ .
\end{align*}%
\newline
Furthermore, \cite{wang2016bounded} derived the efficient influence function
for $\tau $: {\footnotesize 
	\begin{equation*}
	\varphi _{eff}(D,Z,X,Y)=\frac{2Z-1}{f_{Z|X}(Z|X)}\frac{1}{\delta ^{D}(X)}%
	\bigg\{Y-D\delta (X)-\mathbb{E}[Y|Z=0,X]+\mathbb{E}[D|Z=0,X]\delta (X)\bigg\}%
	+\delta (X)-\tau \ ,
	\end{equation*}%
} where $f_{Z|X}(Z|X)$ is the conditional probability mass function of $Z$
given $X$. Clearly, the efficient influence function depends on five unknown
functionals: $\delta (X)$, $\delta ^{D}(X)$, $f_{Z|X}$, $p_{0}^{Y}(X)=%
\mathbb{E}[Y|Z=0,X]$ and $p_{0}^{D}(X)=\mathbb{E}[D|Z=0,X]$. They proposed
to parameterize all five functionals, estimate the functionals with
appropriate parametric approaches, and plug the estimated functionals into
the efficient influence function to estimate $\tau $. They established that
their estimator of $\tau $ is consistent and asymptotically normally
distributed if

\begin{itemize}
	\item either $\delta (X)$, $\delta ^{D}(X)$, $p_{0}^{Y}(X)=\mathbb{E}
	[Y|Z=0,X]$ and $p_{0}^{D}(X)=\mathbb{E}[D|Z=0,X]$ are correctly specified
	\item or $\delta ^{D}(X)$ and $f_{Z|X}$ are correctly specified
	\item or $\delta (X)$ and $f_{Z|X}$ are correctly specified,
\end{itemize}
and their estimator attains the semiparametric efficiency bound only when
all five functionals are correctly specified. The main goal of this paper is
to present an alternative, intuitive and easy approach to compute estimator that does
not require parameterization of any of the functionals and is always
consistent and asymptotically normal and attains the semiparametric
efficiency bound.

\section{Point Estimation}\label{sec:estimation}
 To motivate our estimation procedure, we rewrite the treatment effect
 coefficient. Applying the tower law of conditional expectation, we obtain: 
 \begin{align}
 \tau =& \mathbb{E}\left[ \frac{\delta ^{Y}(X)}{\delta ^{D}(X)}\right] =%
 \mathbb{E}\left[ \frac{\mathbb{E}[Y|Z=1,X]}{\delta ^{D}(X)}-\frac{\mathbb{E}%
 	[Y|Z=0,X]}{\delta ^{D}(X)}\right]  \notag \\
 =& \mathbb{E}\left[ \frac{Z}{f_{Z|X}(1|X)}\cdot \frac{\mathbb{E}[Y|Z=1,X]}{%
 	\delta ^{D}(X)}-\frac{1-Z}{f_{Z|X}(0|X)}\cdot \frac{\mathbb{E}[Y|Z=0,X]}{%
 	\delta ^{D}(X)}\right]  \notag \\
 =& \mathbb{E}\left[ \frac{Z}{f_{Z|X}(1|X)}\cdot \frac{\mathbb{E}[Y|Z,X]}{%
 	\delta ^{D}(X)}-\frac{1-Z}{f_{Z|X}(0|X)}\cdot \frac{\mathbb{E}[Y|Z,X]}{%
 	\delta ^{D}(X)}\right]  \notag \\
 =& \mathbb{E}\left[ \left\{ \frac{2Z-1}{f_{Z|X}(Z|X)}\right\} \frac{Y}{%
 	\delta ^{D}(X)}\right] \ .  \label{def:tau}
 \end{align}%
 The above expression suggests a natural and intuitive plugin estimation,
 with $f_{Z|X}(Z|X)$ and $\delta ^{D}(X)$ replaced by some consistent
 estimates. There are many approaches to estimate these functionals including
 parametric and nonparametric approaches, but as noted by \cite%
 {hirano2003efficient}, not all estimates can lead to efficient estimation of 
 $\tau $. In this paper, we present an intuitive and easy way to compute
 estimates of functionals that ensure efficiency of the plugin estimation of $%
 \tau $. To illustrate our procedure, we notice that the following conditions
 hold for any integrable functions $u_{1}(X)$ and $u_{2}(X)$: 
 \begin{align}
 & \mathbb{E}\left[ \frac{Z}{f_{Z|X}(1|X)}u_{1}(X)\right] =\mathbb{E}%
 [u_{1}(X)]\ =\mathbb{E}\left[ \frac{1-Z}{f_{Z|X}(0|X)}u_{1}(X)\right] ,
 \label{moment:p} \\
 & \mathbb{E}\left[ D\left\{ \frac{2Z-1}{f_{Z|X}(Z|X)}\right\} u_{2}(X)\right]
 =\mathbb{E}\left[ \delta ^{D}(X)u_{2}(X)\right] \ ,  \label{moment:r}
 \end{align}
 and \eqref{moment:p} and \eqref{moment:r} uniquely determine $f_{Z|X}(Z|X)$
 and $\delta ^{D}(X)$. These conditions impose restrictions on the unknown
 functionals and they must be taken into account when estimating those
 functionals. One difficulty with these conditions is that they must be
 imposed in an infinite dimmensional functional space. To overcome this
 difficulty, we propose to impose the conditions on a smaller sieve space.
 Specifically, let $u_{K}(X)=(u_{K,1}(X),\ldots ,u_{K,K}(X))^{\top }$ denote
 a known basis functions that can approximate any suitable function $u(X)$
 arbitrarily well (see \cite{chen2007large} or Appendix \ref{sec:uK} for
 further dicussion). Conditions \eqref{moment:p} and \eqref{moment:r} imply
 for any integers $K_{1}$ and $K_{2}$: 
 \begin{equation}
 \mathbb{E}\left[ \frac{Z}{f_{Z|X}(1|X)}u_{K_{1}}(X)\right] =\mathbb{E}%
 [u_{K_{1}}(X)]=\mathbb{E}\left[ \frac{1-Z}{f_{Z|X}(0|X)}u_{K_{1}}(X)\right]
 \   \label{moment:p_uK}
 \end{equation}%
 and 
 \begin{equation}
 \mathbb{E}\left[ D\left\{ \frac{Z}{f_{Z|X}(1|X)}-\frac{1-Z}{f_{Z|X}(0|X)}%
 \right\} u_{K_{2}}(X)\right] =\mathbb{E}[\delta ^{D}(X)u_{K_{2}}(X)].\ 
 \label{moment:r_uK}
 \end{equation}%
 We shall construct estimates of the functionals by imposing the above
 conditions. To ensure consistency, we shall allow $K_{1}$ and $K_{2}$ to
 increase with sample size at appropriate rates.
 \subsection{Estimation of $f_{Z|X}(Z|X)^{-1}$}
 
 Consider estimation of $f_{Z|X}(Z|X)^{-1}$. An obvious approach is to solve $%
 \left\{ w_{i},i=1,2,...,N\right\} $ from the sample analogue of 
 \eqref{moment:p_uK}: 
 \begin{eqnarray}
 \frac{1}{N}\sum_{i=1}^{N}Z_{i}w_{i}u_{K_{1}}({X}_{i}) &=&\frac{1}{N}%
 \sum_{i=1}^{N}u_{K_{1}}({X}_{i});  \label{sample1} \\
 \frac{1}{N}\sum_{i=1}^{N}(1-Z_{i})w_{i}u_{K_{1}}({X}_{i}) &=&\frac{1}{N}%
 \sum_{i=1}^{N}u_{K_{1}}({X}_{i}).  \label{sample2}
 \end{eqnarray}%
 But there are many solutions and all solutions are consistent estimates of $%
 f_{Z|X}(Z|X)^{-1}$. The question is which solution is the best estimate of $%
 f_{Z|X}(Z|X)^{-1}$ in the sense of ensuring efficient estimation of $\tau $.
 Let $\rho (v)$ denote a strictly increasing and concave function and let $%
 \rho ^{\prime }(v)$ denote its first derivative. Denote 
 \begin{equation*}
 \hat{p}(X_{i})\triangleq \frac{1}{N}\rho ^{\prime }(\hat{\lambda}%
 _{K_{1}}^{\top }u_{K_{1}}(X_{i}))\ ,
 \end{equation*}%
 with $\hat{\lambda}_{K_{1}}\in \mathbb{R}^{K}$ maximizing the following objective function 
 \begin{equation}
 \hat{G}(\lambda )\triangleq \frac{1}{N}\sum_{i=1}^{N}Z_{i}\rho (\lambda
 ^{\top }u_{K_{1}}(X_{i}))-\frac{1}{N}\sum_{i=1}^{N}\lambda ^{\top
 }u_{K_{1}}(X_{i})\ .  \label{E:hatG}
 \end{equation}%
 It is easy to show that $N\hat{p}(X)$ satisfies \eqref{sample1}. Moreover, $N%
 \hat{p}(X)$ can be interpreted as a generalized empirical likelihood
 estimator of $f_{Z|X}(1|X)^{-1}$ (see Appendix \ref{sec:duality}) and hence is the best
 estimate. The fact that $\hat{G}(\lambda )$ is globally concave implies that
 its maximand is easy to compute. \newline
 
 Applying the same idea to \eqref{sample2}, we have 
 \begin{equation*}
 \hat{q}(X_{i})\triangleq \frac{1}{N}\rho ^{\prime }(\hat{\beta}%
 _{K_{1}}^{\top }u_{K_{1}}({X}_{i}))\ ,
 \end{equation*}%
 with $\hat{\beta}_{K_{1}}\in \mathbb{R}^{K_{1}}$ maximizing the following
 globally concave objective function 
 \begin{equation}
 \hat{H}(\beta )\triangleq \frac{1}{N}\sum_{i=1}^{N}(1-Z_{i})\rho (\beta
 ^{\top }u_{K_{1}}(X_{i}))-\frac{1}{N}\sum_{i=1}^{N}\beta ^{\top
 }u_{K_{1}}(X_{i}).  \label{E:hatH}
 \end{equation}%
 Again, $N\hat{q}(X)$ satisfies \eqref{sample2} and can be interpreted as a
 generalized empirical likelihood estimatior of $f_{Z|X}(0|X)^{-1}$. \newline
 
 The $\rho (v)$ function can be any increasing and strictly concave function.
 Some examples include $\rho (v)=-\exp (-v)$ for the exponential tilting %
 \citep{kitamura1997information, imbens1998information}, $\rho (v)=\log (1+v)$
 for the empirical likelihood \citep{owen1988empirical, qin1994empirical}, $%
 \rho (v)=-(1-v)^{2}/2$ for the continuous updating of the generalized method
 of moments \citep{hansen1982large, hansen1996finite} and $\rho (v)=v-\exp
 (-v)$ for the inverse logistic.
 
 \subsection{Estimation of $\protect\delta^D(X)$ and $\protect\tau$}
 
 Having estimated $f_{Z|X}(Z|X)^{-1}$, we now apply the same principle to
 estimate $\delta _{D}(X)$. But there is one difference. Here $\delta
 _{D}(X)\in \lbrack -1,1]$ and the $\rho (v)$ function is not suitable. We
 shall use the following strictly convex function 
 \begin{equation*}
 f(x)=\log (e^{x}+e^{-x})
 \end{equation*}%
 whose derivative is the tanh function $f^{\prime }(x)=\frac{e^{x}-e^{-x}}{%
 	e^{x}+e^{-x}}$ with range $[-1,1]$. We estimate $\delta ^{D}(X)$ by 
 \begin{equation*}
 \hat{\delta}^{D}(X)=f^{\prime }(\hat{\gamma}_{K_{2}}^{\top }u_{K_{2}}(X)),
 \end{equation*}%
 with $\hat{\gamma}_{K_{2}}\in \mathbb{R}^{K_{2}}$ maximizing the following
 globally concave function
 
 \begin{equation*}
 \hat{F}({\gamma })=\frac{1}{N}\sum_{i=1}^{N}D_{i}\{Z_{i}N\hat{p}%
 (X_{i})-(1-Z_{i})N\hat{q}(X_{i})\}\cdot \gamma ^{\top }u_{K_{2}}(X_{i})-%
 \frac{1}{N}\sum_{i=1}^{N}f({\gamma }^{\top }u_{K_{2}}(X_{i})).
 \end{equation*}%
 Again, $\hat{\delta}^{D}(X)$ can be interpreted as a generalized empirial
 likelihood estimator and hence is the best estimate. \newline
 
 Finally, the plugin estimator of $\tau $ is given by 
 \begin{equation*}
 \widehat{\tau }=\sum_{i=1}^{N}\left\{ Z_{i}\hat{p}(X_{i})-(1-Z_{i})\hat{q}%
 (X_{i})\right\} Y_{i}/\hat{\delta}^{D}(X_{i}).
 \end{equation*}
 
 \subsection{Large Sample Properties}
 
To establish the large sample properties of $\widehat{\tau }$, we shall
impose the following assumptions:

\begin{assumption} \label{as:EY2}
	$\mathbb{E}\left[\frac{1}{\delta^D(X)^2}\right] < \infty$ and 	$\mathbb{E}\left[\frac{Y^2}{\delta^D(X)^4}\right] < \infty$.
\end{assumption}

\begin{assumption} \label{as:suppX}
	The support $\mathcal{X}$ of $r$-dimensional covariate $X$ is a Cartesian product of $r$ compact intervals.
\end{assumption}

\begin{assumption}\label{as:f&eta}
	We assume that there exist three positive constants $\infty>\eta_1>\eta_2>1>\eta_3>0$ such that	
	$$\eta_2\leq f^{-1}_{Z|X}(z|x)\leq \eta_1\ \  \text{and} \ \ -\eta_3\leq \delta^D(x)\leq \eta_3\ ,\quad  \forall (z,x)\in \{0,1\}\times \mathcal{X}\ .$$
\end{assumption}



 \begin{assumption}\label{as:approximation}
 	There are  $\lambda_K$, $\beta_K$, $\gamma_K$, $\psi_{1K}$, $\psi_{0K}$,  $\phi_{1K}$ and $\phi_{0K}$ in $\mathbb{R}^K$ and $\alpha>0$ such that{\footnotesize
 		\begin{align*}
 		&\sup_{x\in\mathcal{X}}\left|(\rho')^{-1}\left(\frac{1}{f_{Z|X}(1|x)}\right)-\lambda_K^\top u_K(x)\right|=O(K^{-\alpha})\ ,\ \sup_{x\in\mathcal{X}}\left|(\rho')^{-1}\left(\frac{1}{f_{Z|X}(0|x)}\right)-\beta_K^\top u_K(x)\right|=O(K^{-\alpha})\ ,\\
 		&\sup_{x\in\mathcal{X}}\left|(f')^{-1}\left(\delta^D(x)\right)-\gamma_K^\top u_K(x)\right|=O(K^{-\alpha})\ , \\
 		& \sup_{x\in\mathcal{X}}\left|\frac{p_1^Y(x)}{\delta^D(x)}-\psi_{1K}^\top u_K(x)\right|=O(K^{-\alpha})\ , \ \sup_{x\in\mathcal{X}}\left|\frac{p_0^Y(x)}{\delta^D(x)}-\psi_{0K}^\top u_K(x)\right|=O(K^{-\alpha})\ ,\\
 		& \sup_{x\in\mathcal{X}}\left|\frac{p_1^Y(x)}{\delta^D(x)^2}-\phi_{1K}^\top u_K(x)\right|=O(K^{-\alpha})\ , \ \sup_{x\in\mathcal{X}}\left|\frac{p_0^Y(x)}{\delta^D(x)^2}-\phi_{0K}^\top u_K(x)\right|=O(K^{-\alpha})\ ,
 		\end{align*}}
 	as $K\to \infty$, where $p_z^Y(x)=\mathbb{E}[Y|Z=z,X=x]$ for $z\in\{0,1\}$.
 \end{assumption}

\begin{assumption} \label{as:K&N}
	$K_1\asymp K_2 \asymp K\in \mathbb{N}$, $\zeta(K)^4K^3/N\to 0$ and $\sqrt{N}K^{-\alpha}\to 0$, where $\zeta(K)=\sup_{x\in\mathcal{X}}\|u_K(x)\|$ and $\|\cdot\|$ is the usual Frobenius norm defined by $\|A\|=\sqrt{\tr(AA^\top)}$ for any matrix $A$.
\end{assumption}

\begin{assumption} \label{as:rho}
	$\rho$ is a strictly concave function defined on $\mathbb{R}$, i.e. $ \rho''(\gamma) < 0, ~ \forall \gamma \in \mathbb{R}$, and the range of $\rho'$ contains $[\eta_2,\eta_1]$.
\end{assumption}

Assumption \ref{as:EY2} ensures the asymptotic variance to be bounded.
Assumption \ref{as:suppX} restricts the covariates to be bounded. This
condition, though restrictive, is commonly imposed in the nonparametric
regression literature. Assumption \ref{as:f&eta} requires the probability
function to be bounded away from 0 and 1. Condition of this sort is familiar
in the literature.  Assumption \ref{as:approximation}
is needed to control for the approximation bias, and they
are commonly imposed in the nonparametric literature. Assumption \ref{as:K&N}
imposes restrictions on the smoothing parameter so that the proposed
estimator of ATE is root-N consistent. This condition, however, is
practically unhelpful. We shall present a data driven approach to determine $%
K_{1}$ and $K_{2}$. Assumption \ref{as:rho} is a mild restriction on $\rho $
and is satisfied by all important special cases considered in the
literature.\\
 
 Under the above assumptions, the following theorem establishes the
 consistency, asymptotic normality and the semiparmetric efficiency of $\hat{%
 	\tau}$. 
 \begin{theorem}\label{thm:eff}
 	Suppose that the average treatment effects  is identified in \eqref{id:tau},  under Assumptions \ref{as:EY2}-\ref{as:rho}, we have
 	\begin{enumerate}
 		\item $\hat{\tau}\xrightarrow{p}\tau$;
 		\item $\sqrt{N}(\hat{\tau}-\tau)\xrightarrow{d}N(0,V_{eff})$,
 	\end{enumerate}
 	where 	$V_{eff}=\mathbb{E}\left[\varphi_{eff}(D,Z,X,Y)^2\right]$ is the efficient variance bound  developed in \cite{wang2016bounded}.
 \end{theorem}Sketched proof can be found in Appendix \ref{sec:tech_proofs}
 and detailed proofs are provided in the supplementary material.
 
 \section{Variance Estimation}
 
 \label{sec:variance} To conduct the statistical inference on $\tau $, we
 need a consistent estimator of the asymptotic variance of $\widehat{\tau }$.
 Note that the asymptotic varance of $\widehat{\tau }$, {\small \ 
 	\begin{equation*}
 	\mathbb{E}\left[ \left( \frac{2Z-1}{f_{Z|X}(Z|X)}\frac{1}{\delta ^{D}(X)}%
 	\bigg\{Y-D\delta (X)-\mathbb{E}[Y|Z=0,X]+\mathbb{E}[D|Z=0,X]\delta (X)\bigg\}%
 	+\delta (X)-\tau \right) ^{2}\right] \ ,
 	\end{equation*}%
 } depends on five unknown functionals. Direct estimation of the variance
 requires replacing the five unknown functionals with consistent estimates.
 In this section, we present an alternative estimation that does not require
 estimation of those functionals.\newline
 
 To illustrate the idea, we denote:{\footnotesize 
 	\begin{align*}
 	& g_{1}(Z,{X};\lambda )\triangleq Z\rho^{\prime} \left(\lambda^{\top }u_{K_{1}}({X}%
 	)\right)u_{K_{1}}({X})-u_{K_{1}}({X})\ , \\
 	& g_{2}(Z,{X};\beta )\triangleq (1-Z)\rho^{\prime} \left(\beta^{\top }u_{K_{1}}({X}%
 	)\right)u_{K_{1}}({X})-u_{K_{1}}({X})\ , \\
 	& g_{3}(Z,D,{X};\lambda ,\beta ,\gamma )\triangleq D\left\{ Z\cdot \rho
 	^{\prime }\left( \lambda ^{\top }u_{K_{1}}(X)\right) -(1-Z)\cdot \rho
 	^{\prime }\left( \beta ^{\top }u_{K_{1}}(X)\right) \right\} u_{K_{2}}({X}%
 	)-f^{\prime}\left(\gamma^\top u_{K_{2}}({X})\right)u_{K_{2}}({X})\ , \\
 	& g_{4}(Z,D,X,Y;\lambda ,\beta ,\gamma ,\tau )\triangleq \left\{ Z\cdot \rho
 	^{\prime }\left( \lambda ^{\top }u_{K_{1}}(X)\right) -(1-Z)\cdot \rho
 	^{\prime }\left( \beta ^{\top }u_{K_{1}}(X)\right) \right\} Y/f^{\prime}\left(\gamma^\top u_{K_{2}}(X)\right)-\tau \ ,
 	\end{align*}%
 } and 
 \begin{equation*}
 g(Z,D,X,Y;\theta )\triangleq \left( 
 \begin{array}{c}
 g_{1}(Z,{X};\lambda ) \\ 
 g_{2}(Z,{X};\beta ) \\ 
 g_{3}(Z,D,{X};\lambda ,\beta ,\gamma ) \\ 
 g_{4}(Z,D,X,Y;\lambda ,\beta ,\gamma ,\tau )%
 \end{array}
 \right) 
 \end{equation*}
 with $\theta \triangleq (\lambda ,\beta ,\gamma ,\tau )^{\top }$. Let $\hat{%
 	\theta}\triangleq (\hat{\lambda}_{K_{1}},\hat{\beta}_{K_{1}},\hat{\gamma}%
 _{K_{2}},\hat{\tau})^{\top }$ and ${\theta }^{\ast }\triangleq ({\lambda }
 _{K_{1}}^{\ast },{\beta }_{K_{1}}^{\ast },{\gamma }_{K_{2}}^{\ast },{\tau }
 )^{\top }$. Then $\hat{\theta}$ is the moment estimator solving the
 following moment condition: 
 \begin{equation}
 \frac{1}{N}\sum_{i=1}^{N}g(Z_{i},D_{i},X_{i},Y_{i};\hat{\theta})=0.
 \label{moment}
 \end{equation}
 Applying Mean Value Theorem, we obtain 
 \begin{equation}
 0=\frac{1}{N}\sum_{i=1}^{N}g(Z_{i},D_{i},X_{i},Y_{i};\theta ^{\ast })+\frac{1%
 }{N}\sum_{i=1}^{N}\frac{\partial g(Z_{i},D_{i},X_{i},Y_{i};\tilde{\theta})}{%
 	\partial \theta }(\hat{\theta}-\theta ^{\ast })  \label{eq:empirical}
 \end{equation}
 where $\tilde{\theta}=(\tilde{\lambda}_{K_{1}},\tilde{\beta}_{K_{1}},\tilde{
 	\gamma}_{K_{2}},\tilde{\tau})^{\top}$ lies on the line joining $\hat{\theta}$
 and $\theta ^{\ast }$. We show in the supplemental material that 
 \begin{equation}
 \frac{1}{N}\sum_{i=1}^{N}\frac{\partial g(Z_{i},D_{i},X_{i},Y_{i};\tilde{%
 		\theta})}{\partial \theta }=\mathbb{E}\left[ \frac{\partial g(Z,D,X,Y;\theta
 	^{\ast })}{\partial \theta }\right] +o_{p}(1)  \label{eq:op}
 \end{equation}%
 Note that 
 \begin{align}
 \label{eq:tauhat-tau}
 \hat{\tau}-\tau =\mathbf{e}_{2K_{1}+K_{2}+1}^{\top }(\hat{\theta}-\theta
 ^{\ast })\ ,
 \end{align}%
 where $\mathbf{e}_{2K_{1}+K_{2}+1}$ is a $(2K_{1}+K_{2}+1)$-dimensional
 column vector whose last element is $1$ and other components are all of $0$%
 's. \newline
 
 Combining \eqref{eq:empirical}, \eqref{eq:op} and \eqref{eq:tauhat-tau}, we
 obtain 
 \begin{equation*}
 \sqrt{N}(\hat{\tau}-\tau )=-\mathbf{e}_{2K_{1}+K_{2}+1}^{\top }\left\{ 
 \mathbb{E}\left[ \frac{\partial g(Z,D,X,Y;\theta ^{\ast })}{\partial \theta }
 \right] +o_{p}(1)\right\} ^{-1}\frac{1}{\sqrt{N}}%
 \sum_{i=1}^{N}g(Z_{i},D_{i},X_{i},Y_{i};\theta ^{\ast })\ ,
 \end{equation*}
 which in turn implies 
 \begin{equation*}
 V_{eff}=\lim_{N\rightarrow \infty }Var(\sqrt{N}(\hat{\tau}-\tau
 ))=\lim_{N\rightarrow \infty }\mathbf{e}_{2K_{1}+K_{2}+1}^{\top }\left\{
 L\cdot \Omega \cdot (L^{-1})^{\top }\right\} \mathbf{e}_{2K_{1}+K_{2}+1}\ .
 \end{equation*}%
 where 
 \begin{align*}
 & L=\mathbb{E}\left[ \frac{\partial g(Z,D,X,Y;\theta ^{\ast })}{\partial
 	\theta }\right] \ , \\
 & \Omega =\mathbb{E}\left[ g(Z,D,X,Y;\theta ^{\ast })g(Z,D,X,Y;\theta ^{\ast
 })^{\top }\right] \ .
 \end{align*}%
 Therefore, we can define the sandwich estimator for the efficient variance $%
 V_{eff}$ by 
 \begin{equation*}
 \hat{V}=\mathbf{e}_{2K_{1}+K_{2}+1}^{\top }\left\{ \hat{L}^{-1}\cdot \hat{%
 	\Omega}\cdot (\hat{L}^{-1})^{\top }\right\} \mathbf{e}_{2K_{1}+K_{2}+1}\ ,
 \end{equation*}%
 where 
 \begin{align*}
 & \hat{L}=\frac{1}{N}\sum_{i=1}^{N}\frac{\partial g(Z_{i},D_{i},X_{i},Y_{i};%
 	\hat{\theta})}{\partial \theta }; \\
 & \hat{\Omega}=\frac{1}{N}\sum_{i=1}^{N}g(Z_{i},D_{i},X_{i},Y_{i};\hat{\theta%
 })g(Z_{i},D_{i},X_{i},Y_{i};\hat{\theta})^{\top }.
 \end{align*}%
 \begin{theorem}
 	Under Assumptions \ref{as:EY2}-\ref{as:rho}, $\hat{V}$ is a consistent estimator for the asymptotic variance $V_{eff}$.
 \end{theorem}

 \section{Selection of Tuning Parameters}
 
 \label{sec:K} The large sample properties of the proposed estimator permit a
 wide range of values of $K_{1}$ and $K_{2}$. This presents a dilemma for
 applied researchers who have only one finite sample and would like to have
 some guidance on the selection of smoothing parameters. In this section, we
 present a data-driven approach to select $K_{1}$ and $K_{2}$. Notice that $%
 f_{Z|X}(1|X)^{-1}$, $f_{Z|X}(0|X)^{-1}$ and $\delta ^{D}(X)$ satisfy the
 following regression equations: 
 \begin{align*}
 & \mathbb{E}\left[ Zf_{Z|X}(1|X)^{-1}\bigg|X\right] =1\ , \\
 & \mathbb{E}\left[ (1-Z)f_{Z|X}(0|X)^{-1}\bigg|X\right] =1\ , \\
 & \mathbb{E}\left[ D\left\{
 Zf_{Z|X}(1|X)^{-1}-(1-Z)f_{Z|X}(0|X)^{-1}\right\} \bigg|X\right] =\delta
 ^{D}(X)\ .
 \end{align*}%
 Since $N\hat{p}(X)$, $N\hat{q}(X)$ and $\hat{\delta}^{D}(X)$ are consistent
 estimators of $f_{Z|X}(1|X)^{-1}$, $f_{Z|X}(0|X)^{-1}$ and $\delta ^{D}(X)$
 respectively, the mean-squared-error (MSE) of the nuisance parameters $(\hat{%
 	\lambda}_{K_{1}},\hat{\beta}_{K_{1}})$ and $\hat{\gamma}_{K_{2}}$ are
 defined by 
 \begin{align*}
 MSE_{1}(K_{1})=& \sum_{i=1}^{N}\left\{ Z_{i}N\hat{p}(X_{i})-1\right\}
 ^{2}+\sum_{i=1}^{N}\left\{ (1-Z_{i})N\hat{q}(X_{i})-1\right\} ^{2}\ , \\
 MSE_{2}(K_{1},K_{2})=& \sum_{i=1}^{N}\left\{ D_{i}\left\{ Z_{i}N\hat{p}%
 (X_{i})-(1-Z_{i})N\hat{q}(X_{i})\right\} -\hat{\delta}^{D}(X_{i})\right\}
 ^{2}\ .
 \end{align*}%
 The smoothing parameters $K_{1}$ and $K_{2}$ shall be chosen to minimize $%
 MSE_{1}$ and $MSE_{2}$. Specifically, denote the upper bounds of $K_{1}$ and 
 $K_{2}$ by $\bar{K}_{1}$ and $\bar{K}_{2}$ (e.g. $\bar{K}_{1}=\bar{K}_{2}=5$
 in our simulation studies). The data-driven $K_{1}$ and $K_{2}$ are given by 
 \begin{align*}
 & \hat{K}_{1}=\arg \min_{K_{1}\in \{1,...,\bar{K}_{1}\}}MSE_{1}(K_{1})\ , \\
 & \hat{K}_{2}=\arg \min_{K_{2}\in \{1,...,\bar{K}_{2}\}}MSE_{2}(\hat{K}%
 _{1},K_{2})\ .
 \end{align*}
 
 \section{ Simulation Studies \label{sec:simulation}}
 In this section, we conduct a small scale simulation
study to evaluate the finite sample performance of the proposed estimator.
To evaluate the performance of our estimator against the existing
alternatives, particularly the estimators proposed by \cite{wang2016bounded}%
, we adopt the exact same design (i.e., the same data generating processes
(DGP)). In each Monte Carlo run, we generate sample of data from DGP for two
sizes: $N=500$ and $N=1000$ respectively, and from each sample we compute
our estimator and other existing estimators. We then repeat the Monte Carlo
runs for $500$ times. \newline

The observed baseline covariates are $X=(1,X_{2})$, where $X$ include an
intercept term and a continuous random variable $X_{2}$ uniformly
distributed on the interval $(-1,-0.5)\cup (0.5,1)$. The unmeasured
confounder $U$ is a Bernoulli random variable with mean 0.5. The
instrumental variable $Z$, treatment variable $D$ and outcomes variable $%
Y\in \{0,1\}$ are generated according to the simulation design of \cite%
{wang2016bounded}. The true value of the average treatment effect is $\tau
=0.087$.\newline

We compute the proposed estimator (cbe), the naive estimator, the multiply
robust estimator (mr) and the bounded multiply robust estimator (b-mr)
proposed by \cite{wang2016bounded}. Details of calculations are given below.

\begin{enumerate}
	\item the proposed estimator (cbe) is computed with $\rho (v)=\log (1+v)$;
	
	\item the naive estimator is computed by the difference of group means
	between treatment and control groups;
	
	\item the multiply robust estimator (mr) and the bounded multiply robust
	estimator (b-mr) are computed by the procedures proposed by \cite%
	{wang2016bounded}.
\end{enumerate}

The multiply robust estimator (mr) and the bounded multiply robust estimator
(b-mr) proposed by \cite{wang2016bounded} depend on parameterization of five
unknown functionals. In their paper they considered several models, denoted
by $\mathcal{M}_{1}$, $\mathcal{M}_{2}$ and $\mathcal{M}_{3}$ (see \cite%
{wang2016bounded} for a detailed discussion of the model specification).
Following \cite{wang2016bounded}, we consider scenarios where some or all
functionals are misspecified. \newline
\baselineskip0.6cm 
\begin{table}[tph]
	\caption{Simulation results of estimated average treatment effects }
	\label{table:sim_results_clayton_high}
	\begin{center}
		{\fontsize{9pt}{12.8pt} \selectfont 
			\setlength{\tabcolsep}{12mm}{\ 
				\begin{tabular}{lccc}
					\hline\hline
					\multicolumn{4}{c}{$N = 500$} \\ \hline
					Estimators & Bias & Stdev & RMSE \\ 
					Naive & -0.057 & 0.045 & 0.073 \\ 
					mr(All) & 0.003 & 0.139 & 0.139 \\ 
					mr($\mathcal{M}_1$) & 0.004 & 0.139 & 0.139 \\ 
					mr($\mathcal{M}_2$) & -0.004 & 0.163 & 0.163 \\ 
					mr($\mathcal{M}_3$) & -30.973 & 883.036 & 884.579 \\ 
					mr(None) & -13.887 & 419.412 & 419.648 \\ 
					b-mr(All) & 0.006 & 0.145 & 0.145 \\ 
					b-mr($\mathcal{M}_1$) & -0.015 & 0.163 & 0.164 \\ 
					b-mr($\mathcal{M}_2$) & -0.010 & 0.207 & 0.207 \\ 
					b-mr($\mathcal{M}_3$) & 0.008 & 0.142 & 0.143 \\ 
					mr(None) & -0.137 & 0.648 & 0.663 \\ 
					cbe & 0.003 & 0.152 & 0.152 \\ \hline\hline
					\multicolumn{4}{c}{$N = 1000$} \\ \hline
					Estimators & Bias & Stdev & RMSE \\ 
					Naive & -0.056 & 0.031 & 0.064 \\ 
					mr(All) & -0.002 & 0.102 & 0.102 \\ 
					mr($\mathcal{M}_1$) & -0.0005 & 0.102 & 0.102 \\ 
					mr($\mathcal{M}_2$) & -0.011 & 0.121 & 0.121 \\ 
					mr($\mathcal{M}_3$) & -94.930 & 1737.95 & 1740.541 \\ 
					mr(None) & 9.708 & 240.259 & 240.455 \\ 
					b-mr(All) & 0.003 & 0.104 & 0.104 \\ 
					b-mr($\mathcal{M}_1$) & -0.021 & 0.134 & 0.136 \\ 
					b-mr($\mathcal{M}_2$) & -0.008 & 0.141 & 0.141 \\ 
					b-mr($\mathcal{M}_3$) & 0.002 & 0.103 & 0.103 \\ 
					b-mr(None) & 0.224 & 0.638 & 0.676 \\ 
					cbe & 0.004 & 0.110 & 0.110 \\ \hline\hline
				\end{tabular}
		} }
	\end{center}
	\par
	{\fontsize{8.5pt}{11.6pt} \selectfont 
		The true value for of the average tratment effects is 0.087. Bias, standard
		deviation (Stdev), root mean squared error (RMSE) of each estimator after $J
		= 500$ Monte Carlo trials are reported. All: all of the three models $%
		\mathcal{M}_1,\mathcal{M}_1,\mathcal{M}_3$ are correctly specified; $%
		\mathcal{M}_1$: only the model $\mathcal{M}_1$ is correctly specified; $%
		\mathcal{M}_2$: only the model $\mathcal{M}_2$ is correctly specified; $%
		\mathcal{M}_3$: only the model $\mathcal{M}_3$ is correctly specified; None:
		all of the models are misspecified. }
\end{table}

\begin{table}[tph]
	\caption{Simulation results of estimated efficient deviation }
	\label{table:3}
	\begin{center}
		{\fontsize{9pt}{12.8pt} \selectfont 
			\setlength{\tabcolsep}{15mm}{\ 
				\begin{tabular}{ccc}
					\hline\hline
					\multicolumn{3}{c}{$N = 500$} \\ \hline
					Methods & Situation & Deviation Estimate \\ 
					& All & 3.04 \\ 
					& $\mathcal{M}_1 $ & 3.19 \\ 
					mr & $\mathcal{M}_2$ & 3.22 \\ 
					& $\mathcal{M}_3$ & 2260.0 \\ 
					& None & 3596.7 \\ 
					& All & 3.04 \\ 
					& $\mathcal{M}_1$ & 3.19 \\ 
					b-mr & $\mathcal{M}_2$ & 3.22 \\ 
					& $\mathcal{M}_3$ & 2078.0 \\ 
					& None & 3572.2 \\ 
					cbe & ---- & 3.41 \\ \hline\hline
					\multicolumn{3}{c}{$N = 1000$} \\ \hline
					Methods & Situation & Deviation Estimate \\ 
					& All & 3.04 \\ 
					& $\mathcal{M}_1 $ & 3.20 \\ 
					mr & $\mathcal{M}_2$ & 3.22 \\ 
					& $\mathcal{M}_3$ & 2291.9 \\ 
					& None & 1363.0 \\ 
					& All & 3.04 \\ 
					& $\mathcal{M}_1 $ & 3.20 \\ 
					b-mr & $\mathcal{M}_2$ & 3.23 \\ 
					& $\mathcal{M}_3$ & 1491.8 \\ 
					& None & 1341.8 \\ 
					cbe & ---- & 3.36 \\ \hline\hline
				\end{tabular}
		} }
	\end{center}
	\par
	{\fontsize{8.5pt}{11.6pt} \selectfont 
		The true value of efficient deviation is 3.04. All: all of the three models $%
		\mathcal{M}_1,\mathcal{M}_1,\mathcal{M}_3$ are correctly specified; $%
		\mathcal{M}_1$: only the model $\mathcal{M}_1$ is correctly specified; $%
		\mathcal{M}_2$: only the model $\mathcal{M}_2$ is correctly specified; $%
		\mathcal{M}_3$: only the model $\mathcal{M}_3$ is
		correctly specified; None: all of the models are misspecified. }
\end{table}

\begin{figure}[tph]
	\caption{Histogram of $K_{1}$ \& $K_{2}$}
	\begin{center}
	\begin{subfigure}[t]{150mm}
		\centering
	   \includegraphics[height=6cm,width=12cm, scale=1]{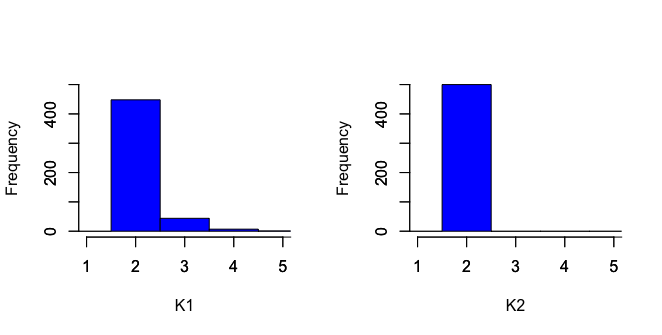}
			\caption{n=500}
	\end{subfigure}	
	\par
	\begin{subfigure}[t]{150mm}
		\centering
	   \includegraphics[height=6cm,width=12cm, scale=1]{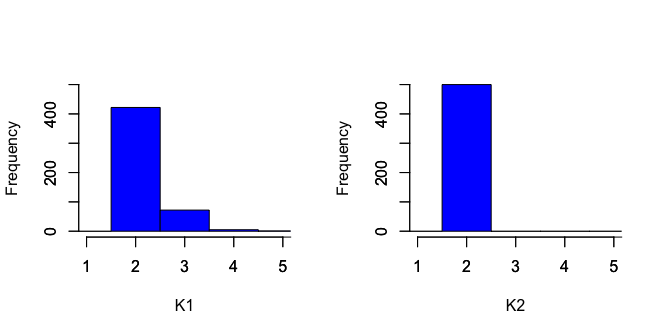}
	  \caption{n=1000}
	\end{subfigure}
\end{center}
\end{figure}

Table 1 reports the bias, standard deviation (Stdev), and the root mean
square error (RMSE) of $\widehat{\tau }$ from the 500 Monte Carlo runs. In
each Monte Carlo run, we use the data driven approach to select $K_{1}$ and $%
K_{2},$ and their histograms are depicted in Figure 1. The estimated
asymptotic variances are reported in Table 2. \newline

Glancing at these tables, we have the following observations:

\begin{enumerate}
	\item The naive estimator has large bias. This is not surprising since it
	ignores the confounding effect.
	
	\item The multiple robust estimators (mr) of \cite{wang2016bounded} has huge
	bias when some functionals are misspecified.
	
	\item The bounded multiple robust estimator (b-mr) of \cite{wang2016bounded}
	is more robust than mr-estimator, but it still has a significant bias if
	some functionals are misspecified. And the bias does not valish as the
	sample size increases. Moreover, if all functionals are misspecified, the
	bias of b-mr estimator is substantially large.
	
	\item The proposed estimator (cbe) is unbiased for both $N=500$ and $N=1000$%
	. Its performance (Bias, Stdev, RMSE) is comparable to \cite{wang2016bounded}
	's estimator when all functionals are correctly parameterized.
	
	\item In variance estimation, both the multiple robust estimator (mr) and
	the bounded multiple robust estimator (b-mr) have large biases when some
	functionals are misspecified. In contrast, the proposed variance estimator
	is consistent.
	
	\item The histograms in Figure 1 reveal that for both $N=500$ and $N=1000$, $%
	K_{1}=2$ and $K_{2}=2$ are most preferred, suggesting that the growing rate
	of $K_{1}$ and $K_{2}$ is slow, an observation consistent with Assumption %
	\ref{as:K&N}.
\end{enumerate}

Overall, the simulation results show that the proposed estimator
out-performs the existing estimators.

\section{Concluding Remarks}

\label{sec:discussion} Most of the existing treatment effect literature on
observational data assume that all confounders are observed and available to
researchers. In applications, it is often the case that some confounders are
not observed or not available. \cite{wang2016bounded} studied identification
and estimation of the average treament effect when some confounders are not
observed. They propose to parameterize five unknown functionals and show
that their estimation is consistent when certain functionals are correctly
specified and is efficient when all functionals are correctly specified.
This paper proposes an alternative estimation. Unlike \cite{wang2016bounded}%
, the proposed estimation does not parameterize any of the functionals and
is always consistent. Moreover, the proposed estimator attains the
semiparametric efficiency bound. A simple asymptotic variance estimator is
presented, and a small scale simulation study suggests the practicality of
the proposed procedure.\newline

Our procedure only applies to the binary treatment with unmeasured
confounders. However, other forms of treatment, such as multiple valued or
continuous treatment, may arise in applications. Extension of the proposed
methodology to those forms of treatment with unmeasured confounders is
certainly of great interest. This extension shall be pursued in a future
project.

\singlespacing

\bibliographystyle{econometrica}
\bibliography{Semiparametric}

\appendix

\section{Appendix }\label{sec:appendix}

	\subsection{Discussion on $u_K$}\label{sec:uK}	
To construct our estimator, we need to specify  the sieve basis $u_K(X)$. Although the approximation theory is derived for general sequences of sieve basis, the most common class of functions
are power series and splines. In particular, we can approximate any function $f:\mathbb{R}^{r}\to \mathbb{R}$
by $\tilde{\gamma}_{K}^{\top}\tilde{u}_{K}(x)$, where $\tilde{u}_K(x)$ is a prespecified sieve basis. Because $\tilde{\gamma}^{\top}_{K}\tilde{u}_{K}(x)=\tilde{\gamma}_{K}^{\top}A_{K\times K}^{-1}A_{K\times K}\tilde{u}_{K}(x)$, we can also use  $u_{K}(x)=A_{K\times K}\tilde{u}_{K}(x)$  as the new basis for approximation. By choosing $A_{K\times K}$ appropriately we obtain a system of orthonormal  basis (with respect to some weights). In particular, we choose $A_{K\times K}$ so that 
\begin{align}\label{eq:orthonormal_basis}
\mathbb{E}\left[u_{K}(X)u_{K}^{\top}(X)\right]=I_{K\times K}  \ . 
\end{align}
We define the usual Frobenius norm $\|A\| \triangleq \sqrt{\tr(AA^{\top})}$ for any matrix $A$. Define 
\begin{align}
\zeta(K) \triangleq \sup_{x\in\mathcal{X}}\|u_{K}(x)\|\ .
\end{align}
In general, this bound depends on the array of basis that is used. \cite{Newey94, Newey97} showed that
\begin{enumerate}
\item  for  power series:  there exists a universal constant $C_0 >0$ such that $\zeta(K) \leq C_0K$;
\item for regression splines:  there exists a universal constant $C_0 >0$ such that $\zeta(K) \leq C_0\sqrt{K}$.
\end{enumerate}

\subsection{Duality of Constrained Optimization}
\label{sec:duality} Let $L(v,v_0)$ be a distance measure that is
continuously differentiable in $v\in \mathbb{R}$, non-negative, strictly
convex in $v$ and $L(v_0,v_0)=0$. The general idea of calibration is to
minimize the aggregate distance between the final weights to a given vector
of design weights subject to moment constraints. Being motivated by %
\eqref{moment:p_uK}, we consider to construct the calibration weights $%
\{w_i\}_{i=1}^N$ by solving the following constrained optimization problem: 
\begin{align}  \label{E:cm}
\left\{ 
\begin{array}{ll}
& \qquad \qquad \qquad \qquad \text{Minimize} ~~~ \sum_{i=1}^N L(w_i,1) \ ,
\\[3mm] 
& \text{subject to}~~\frac{1}{N}\sum_{i=1}^NZ_iw_iu_{K_1}({X}_i)=\frac{1}{N}%
\sum_{i=1}^Nu_{K_1}({X}_i)=\frac{1}{N}\sum_{i=1}^N(1-Z_i)w_iu_{K_1}({X}_i)%
\end{array}
\right. \ ,
\end{align}
where $K_1\to \infty$ as the sample size $N\to \infty$, yet with $K_1/N\to 0$. The constrained optimization problem stated above is equivalent to two
separate constrained optimization problems. 
\begin{align}
&\text{Minimize} ~~~ \sum_{i=1}^N Z_iL(Np_i,1) ~~\text{subject to}%
~~\sum_{i=1}^NZ_ip_iu_{K_1}({X}_i)=\frac{1}{N}\sum_{i=1}^Nu_{K_1}({X}_i)\ ,
\label{E:cm1} \\
&\text{Minimize} ~~~ \sum_{i=1}^N (1-Z_i)L(Nq_i,1) ~~\text{subject to}%
~~\sum_{i=1}^N (1-Z_i)q_iu_{K_1}({X}_i)=\frac{1}{N}\sum_{i=1}^Nu_{K_1}({X}%
_i)\ \ .  \label{E:cm2}
\end{align}
Because the primal problems (\ref
{E:cm1}) and (\ref{E:cm2}) are convex separable programs with linear
constraints, \cite{tseng1987relaxation} showed that the dual problems
are unconstrained convex maximization problems that can be solved by
numerical efficient and stable algorithms. \newline

We show the dual of  (\ref{E:cm1}) is the unconstrained optimization \eqref{E:hatG} by using
the methodology introduced in \cite{tseng1987relaxation}. Let $g(v)=L(1-v,1)$, $g^{\prime
}(v)=\partial g(v)/\partial v$, $E_{K_1\times N}\triangleq
\left(u_{K_1}(X_1),\ldots,u_{K_1}(X_N)\right)$, $s_i\triangleq 1-Z_iNp_i, 
i=1,\ldots,N$, and $\mathbf{s}\triangleq \left(s_1,\ldots,s_N\right)^\top$,
then we can rewrite the problem (\ref{E:cm1}) as 
\begin{equation*}
\min_{\mathbf{s}} \sum_{i=1}^NZ_ig(s_i) ~~~\text{subject to} ~~~~
E_{K_1\times N} \cdot\mathbf{s}=0 \ .
\end{equation*}
For every $j\in \{1,\ldots,N\}$, we define the conjugate convex function %
\citep{tseng1987relaxation} of $Z_jg(\cdot)$ to be 
\begin{align*}
l_{j}(u_j)=&\sup_{s_j}\left\{u_js_j-Z_jg(s_j)\right\}=\sup_{p_j}\left%
\{-Z_jNp_ju_j+u_j-Z_jg(1-Z_jNp_j)\right\} \\
=&\sup_{p_j}\left\{-Z_jNp_ju_j+u_j-Z_jg(1-Np_j)\right\} \\
=&-Z_jNp^*_ju_j+u_j-Z_jg(1-Np^*_j)\ ,
\end{align*}
where the third equality follows by $Zg(1-ZNp_j)=Zg(1-Np_j)$, and $p_j^*$
satisfies the first order condition: 
\begin{align*}
-Z_ju_j=-Z_jg^{\prime }(1-Np_j^*)\Rightarrow p_j^*=\frac{1}{N}%
\left\{1-\left(g^{\prime }\right)^{-1}(u_j)\right\}\ ;
\end{align*}
then we can have 
\begin{align*}
l_{j}(u_j)=&-Z_ju_j\left\{1-\left(g^{\prime
}\right)^{-1}(u_j)\right\}+u_j-Z_jg\left(\left(g^{\prime
}\right)^{-1}(u_j)\right) \\
=&-Z_j\left\{g\left(\left(g^{\prime
}\right)^{-1}(u_j)\right)+u_j-u_j\left(g^{\prime
}\right)^{-1}(u_j)\right\}+u_j \\
=&-Z_j\rho\left(u_j\right)+u_j \ ,
\end{align*}
where 
\begin{equation*}
\rho\left( u\right)\triangleq g\left(\left(g^{\prime
}\right)^{-1}(u)\right)+u-u\left(g^{\prime }\right)^{-1}(u)\ .
\end{equation*}
By \cite{tseng1987relaxation}, the dual problem of (\ref{E:cm1}) is
\begin{align*}
&\min_{\lambda} \sum_{j=1}^N l_j(\lambda^\top E_j)=\min_{\lambda}
\sum_{j=1}^N l_j(\lambda^\top u_{K_1}(X_j)) \\
=&\min_{\lambda} \sum_{j=1}^N\left\{-Z_j\rho\left(\lambda^\top
u_{K_1}(X_j)\right)+\lambda^\top u_{K_1}(X_j)\right\} \\
=&-\max_{\lambda} \sum_{j=1}^N\left\{Z_j\rho\left(\lambda^\top
u_{K_1}(X_j)\right)-\lambda^\top u_{K_1}(X_j)\right\} \\
=&-\max_{\lambda} \hat{G}(\lambda)\ ,
\end{align*}
where $E_j$ is the $j$-th column of $E_{K_1\times N}$, i,e., $%
E_j=u_{K_1}(X_j)$, which is our formulation \eqref{E:hatG}. \newline

Since $L(\cdot)$ is strictly convex, i.e., $L^{\prime \prime }(v)>0$, and $%
g^{\prime \prime }(v)=L^{\prime \prime }(1-v)$, then $g(\cdot)$ is also
strictly convex and $g^{\prime }(\cdot)$ is strictly increasing. Note that 
\begin{align*}
\rho(v) =g((g^{\prime -1}(v))+v-v(g^{\prime -1}(v) 
\Leftrightarrow \rho\left(g^{\prime }(v)\right) =g(v)+g^{\prime
}(v)-vg^{\prime }(v) \ .
\end{align*}
Differentiating $v$ on both sides in above equation yields: 
\begin{align*}
\rho^{\prime }\left(g^{\prime }(v)\right)g^{\prime \prime }(v) =g^{\prime
}(v)+g^{\prime \prime }(v)-g^{\prime }(v)-vg^{\prime \prime
}(v)=(1-v)g^{\prime \prime }(v) \ .
\end{align*}
Since $g^{\prime \prime }(v)>0$, we can have 
\begin{align*}
\rho^{\prime }\left(g^{\prime }(v)\right)=1-v \ ,
\end{align*}
then  we differentiate $v$ on both sides to get $\rho^{\prime \prime
}\left(g^{\prime }(v)\right)g^{\prime \prime }(v)=-1$, which implies 
\begin{equation*}
\rho^{\prime \prime }(v)=-\frac{1}{g^{\prime \prime }\left((g^{\prime
		-1}(v)\right)} <0 \ .
\end{equation*}
Therefore, the convexity of $L(\cdot)$ is equivalent to the concavity of $%
\rho(\cdot)$.

\subsection{Convergence Rates of Estimated Weights}
The following result ensures the consistency of 
$N\hat{p}(X)$, $N\hat{q}(X)$ and $\hat{\delta}^{D}(X)$ as well as their
convergence rates. The proof is presented  in Section 2 of the supplemental material. 
	\begin{prop}\label{prop:hatp-p}
		Under Assumptions \ref{as:suppX}-\ref{as:rho}, we have
		\begin{align*}
		&\sup_{x\in\mathcal{X}}|N\hat{p}(x)-f_{Z|X}(1|x)^{-1}|=O_p\left(\zeta(K)K^{-\alpha}+\zeta(K)\sqrt{\frac{K}{N}}\right)\ , \\
		&\int_{\mathcal{X}}|N\hat{p}(x)-f_{Z|X}(1|x)^{-1}|^2dF_X(x)=O_p\left(K^{-2\alpha}+{\frac{K}{N}}\right)\ , \\
		&\frac{1}{N}\sum_{i=1}^N|N\hat{p}(X_i)-f_{Z|X}(1|X_i)^{-1}|^2=O_p\left(K^{-2\alpha}+{\frac{K}{N}}\right)\ , 
		\end{align*} 
		and
		\begin{align*}
		&\sup_{x\in\mathcal{X}}|N\hat{q}(x)-f_{Z|X}(0|x)^{-1}|=O_p\left(\zeta(K)K^{-\alpha}+\zeta(K)\sqrt{\frac{K}{N}}\right)\ ,\\
		&\int_{\mathcal{X}}|N\hat{q}(x)-f_{Z|X}(0|x)^{-1}|^2dF_X(x)=O_p\left(K^{-2\alpha}+{\frac{K}{N}}\right)\ , \\
		&\frac{1}{N}\sum_{i=1}^N|N\hat{q}(X_i)-f_{Z|X}(0|X_i)^{-1}|^2=O_p\left(K^{-2\alpha}+{\frac{K}{N}}\right)\ ,
		\end{align*}
		and	
		\begin{align*}
		&\sup_{x\in\mathcal{X}}|\hat{\delta}^D(x)-\delta^D(x)|=O_p\left(\zeta(K)K^{-\alpha}+\zeta(K)\sqrt{\frac{K}{N}}\right)\ ,\\
		&\int_{\mathcal{X}}|\hat{\delta}^D(x)-\delta^D(x)|^2dF_X(x)=O_p\left(K^{-2\alpha}+{\frac{K}{N}}\right) \ , \\
		&\frac{1}{N}\sum_{i=1}^N|\hat{\delta}^D(X_i)-\delta^D(X_i)|^2=O_p\left(K^{-2\alpha}+{\frac{K}{N}}\right)\ .
		\end{align*}
	\end{prop}

\subsection{Sketched Proof of Theorem \ref{thm:eff}}

\label{sec:tech_proofs} The detailed proof of Theorem \ref{thm:eff} is given in the
supplementary material. Here we present the outline of whole the proof. By Assumption \ref{as:K&N}, $K_1\asymp K_2\asymp K$, without loss of
generality, we assume that $K_1=K_2=K$%
. We introduce the following notation: let $G^*(\lambda)$, $\lambda_K^*$
and $p^*(X)$ be the theoretical counterparts of $\hat{G}(\lambda)$, $
\hat{\lambda}_K$ and $\hat{p}(X)$ defined by 
\begin{align*}
&G^*(\lambda)=\mathbb{E}[\hat{G}_K(\lambda)]=\mathbb{E}\left[Z\rho^{\prime
}\left(\lambda^\top u_K(X)\right)-\lambda^\top u_K(X)\right]\ , \\
&\lambda_K^*=\arg\max G^*(\lambda)\ , \ {p}^*(X)=\frac{1}{N}\rho'(({\lambda}_K^*)^\top u_K(X))\ .
\end{align*}
We also introduce the following notation: {
	\begin{align*}
	&p^Y_1(X)=\mathbb{E}[Y|Z=1,X]\ ,\ p^Y_0(X)=\mathbb{E}[Y|Z=0,X]\ , \ \delta^Y(X)=p^Y_1(X)-p^Y_0(X)\ ,\\
	&\tilde{\Psi}_K=-\int_{\mathcal{X}}\frac{p_1^Y(x)}{\delta^D(x)}%
	f_{Z|X}(1|x)\rho^{\prime \prime }(\tilde{\lambda}_K^{%
		\top}u_K(x))u_K(x)dF_X(x) \ , \\
	&{\Psi}_K=-\int_{\mathcal{X}}\frac{p_1^Y(x)}{\delta^D(x)}f_{Z|X}(1|x)\rho^{%
		\prime \prime }(({\lambda}^*_K)^{\top}u_K(x))u_K(x)dF_X(x)\ , \\
	&\tilde{\Sigma}_K=\frac{1}{N}\sum_{i=1}^NZ_i\rho^{\prime \prime }(\tilde{%
		\lambda}_K^{\top}u_K(X_i))u_{K}(X_i)u_{K}(X_i)^{\top}\ , \\
	&\Sigma_K=-\mathbb{E}\left[f_{Z|X}(1|X)\rho^{\prime \prime }(({\lambda}%
	_K^*)^{\top}u_K(X))u_{K}(X)u_{K}(X)^{\top}\right]\ , \\
	&\tilde{Q}_K(X)=\tilde{\Psi}_K^{\top}\tilde{\Sigma}_K^{-1}u_K(X)\ , \ {Q}%
	_K(X)={\Psi}_K^{\top}{\Sigma}_K^{-1}u_K(X)\ ,
	\end{align*}}
where $\tilde{\lambda}_K$ lies on the line joining $\hat{\lambda}_K$ and $\lambda_K^*$.  Note that $Q_K(X)$ is the weighted $L^2$
projection of $-p^Y_1(X)/\delta^D(X)$ on the space linearly spanned by $u_K(X)$%
. Note that 
\begin{equation*}
\sqrt{N}(\hat{\tau}-\tau)=\sqrt{N}\sum_{i=1}^NZ_i\hat{p}(X_i)Y_i/\hat{\delta}%
^D(X_i)-\sqrt{N}\sum_{i=1}^N(1-Z_i)\hat{q}(X_i)Y_i/\hat{\delta}^D(X_i)\ .
\end{equation*}
We first derive the influence function of $\sqrt{N}\sum_{i=1}^NZ_i\hat{p}%
(X_i)Y_i/\hat{\delta}^D(X_i)$, and similarly obtain that of $\sqrt{N}%
\sum_{i=1}^N(1-Z_i)\hat{q}(X_i)Y_i/\hat{\delta}^D(X_i)$. We can decompose $%
\sqrt{N}\sum_{i=1}^NZ_i\hat{p}(X_i)Y_i/\hat{\delta}^D(X_i)$ as follows:{\footnotesize
	\begin{align}
	&\sqrt{N}\sum_{i=1}^NZ_i\hat{p}(X_i)Y_i/\hat{\delta}^D(X_i)\notag\\
	=&\frac{1}{\sqrt{N}}\sum_{i=1}^N\frac{Z_i}{\hat{\delta}^D(X_i)}\{N\hat{p}(X_i)-N{p}^*(X_i)\}Y_i -\frac{1}{\sqrt{N}}\sum_{i=1}^N\frac{Z_i}{\delta^D(X_i)}\{N\hat{p}(X_i)-N{p}^*(X_i)\}Y_i \label{eq:delta^-delta_p-p*} \\
	& +\frac{1}{\sqrt{N}}\sum_{i=1}^N\frac{Z_i}{\hat{\delta}^D(X_i)}N{p}^*(X_i)Y_i -\frac{1}{\sqrt{N}}\sum_{i=1}^N\frac{Z_i}{\delta^D(X_i)}N{p}^*(X_i)Y_i \label{eq:delta^-delta} \\
	\label{eq:WK1}&~ + \frac {1} {\sqrt{N}} \sum_{i=1}^N \Bigg\{ \frac{Z_i}{\delta^D(X_i)}\left(N\hat{p}(X_i) - Np^*(X_i)\right)Y_i  - \int_{\mathcal{X}} \frac{p_1^Y(x)f_{Z|X}(1|x)}{\delta^D(x)}(N\hat{p}(X) - Np^*(X))dF_X(x)\Bigg\} \\
	\label{eq:VK} &~ + \frac {1} {\sqrt{N}} \sum_{i=1}^N \Bigg\{\left(Np^*(X_i) - \frac {1} {f_{Z|X}(1|X_i)}\right)\frac{Z_iY_i}{\delta^D(X_i)} - \mathbb{E}\left[\frac{p_1^Y(X)}{\delta^D(X)}f_{Z|X}(1|X)\left(Np^*(X) - \frac {1} {f_{Z|X}(1|X)}\right)\right] \Bigg\} \\
	\label{eq:Lemma1}&~ + \sqrt{N}\mathbb{E}\left[\frac{p_1^Y(X)}{\delta^D(X)}f_{Z|X}(1|X)\left(Np^*(X) - \frac {1} {f_{Z|X}(1|X)}\right)\right]\\
	&~ \label{eq:tauto}+ \sqrt{N}\int_{\mathcal{X}}\frac{p_1^Y(x)}{\delta^D(x)}f_{Z|X}(1|x)(N\hat{p}(X) - Np^*(X))dF_X(x)  - \frac {1} {\sqrt{N}} \sum_{i=1}^N [Z_i \rho'((\lambda_K^*)^{\top}u_K(X_i)) - 1] \tilde{Q}_K(X_i)\\
	&~ \label{eq:Q} + \frac {1} {\sqrt{N}} \sum_{i=1}^N[Z_i\rho'((\lambda_K^*)^{\top}u_K(X_i)) - 1](\tilde{Q}_K(X_i) - Q_K(X_i)) \\
	&~ \label{eq:proj} + \frac {1} {\sqrt{N}} \sum_{i=1}^N \left\{[Z_i\rho'((\lambda_K^*)^{\top}u_K(X_i)) - 1]Q_K(X_i)
	+ \frac {p_1^Y(X_i)} {\delta^D(X_i)}\left(\frac{Z_i}{f_{Z|X}(1|X_i)} - 1\right)\right\} \\
	\label{eq:Normal}&~ + \frac {1} {\sqrt{N}} \sum_{i=1}^N \left\{\frac {Z_iY_i} {f_{Z|X}(1|X_i)\delta^D(X_i)} 
	- \frac {p_1^Y(X_i)} {\delta^D(X_i)}\left(\frac{Z_i}{f_{Z|X}(1|X_i)} - 1\right) \right\}\ .
	\end{align}}
The following lemmas are proved in the supplemental material.
	\begin{lemma}\label{lemma:op(1)}
	Under Assumptions \ref{as:EY2}-\ref{as:rho}, the terms \eqref{eq:delta^-delta_p-p*} \eqref{eq:WK1}, \eqref{eq:VK}, \eqref{eq:Lemma1}, \eqref{eq:tauto}, \eqref{eq:Q} and \eqref{eq:proj} are of $o_p(1)$
	\end{lemma}
	\begin{lemma} \label{lemma:delta^-delta}Under Assumptions \ref{as:EY2}-\ref{as:rho}, \eqref{eq:delta^-delta} has the following equivalent linear expression: 
		\begin{align*}
		\eqref{eq:delta^-delta}=	
		-\frac{1}{\sqrt{N}}\sum_{i=1}^ND_i\cdot\frac{2Z_i-1}{f_{Z|X}(Z_i|X_i)}\cdot \frac{p_1^Y(X_i)}{\delta^D(X_i)^2}+\frac{1}{\sqrt{N}}\sum_{i=1}^N\frac{2Z_i-1}{\delta^D(X_i)^2}\cdot \frac{\mathbb{E}[D_i|Z_i,X_i]}{f_{Z|X}(Z_i|X_i)}p_1^Y(X_i)+o_p(1)\ .
		\end{align*}
	\end{lemma}
	By Lemmas \ref{lemma:op(1)} and \ref{lemma:delta^-delta}, we can obtain that
	\begin{align*}
	&\sqrt{N}\sum_{i=1}^NZ_i\hat{p}(X_i)Y_i/\hat{\delta}^D(X_i)\\
	=&\frac {1} {\sqrt{N}} \sum_{i=1}^N \left\{\frac {Z_iY_i} {f_{Z|X}(1|X_i)\delta^D(X_i)} 
	- \frac {p_1^Y(X_i)} {\delta^D(X_i)}\left(\frac{Z_i}{f_{Z|X}(1|X_i)} - 1\right) \right\}\\
	&-\frac{1}{\sqrt{N}}\sum_{i=1}^ND_i\cdot\frac{2Z_i-1}{f_{Z|X}(Z_i|X_i)}\cdot \frac{p_1^Y(X_i)}{\delta^D(X_i)^2}+\frac{1}{\sqrt{N}}\sum_{i=1}^N\frac{2Z_i-1}{\delta^D(X_i)^2}\cdot \frac{\mathbb{E}[D_i|Z_i,X_i]}{f_{Z|X}(Z_i|X_i)}p_1^Y(X_i)+o_p(1)\ .
	\end{align*}
	Symmetrically, we have
	\begin{align*}
	&\sqrt{N}\sum_{i=1}^N(1-Z_i)\hat{q}(X_i)Y_i/\hat{\delta}^D(X_i)\\
	=&\frac {1} {\sqrt{N}} \sum_{i=1}^N \left\{\frac {(1-Z_i)Y_i} {f_{Z|X}(0|X_i)\delta^D(X_i)} 
	- \frac {p^Y_0(X_i)} {\delta^D(X_i)}\left(\frac{1-Z_i}{f_{Z|X}(0|X_i)} - 1\right) \right\}\\
	&-\frac{1}{\sqrt{N}}\sum_{i=1}^ND_i\cdot\frac{2Z_i-1}{f_{Z|X}(Z_i|X_i)}\cdot \frac{p^Y_0(X_i)}{\delta^D(X_i)^2}+\frac{1}{\sqrt{N}}\sum_{i=1}^N\frac{2Z_i-1}{\delta^D(X_i)^2}\cdot \frac{\mathbb{E}[D_i|Z_i,X_i]}{f_{Z|X}(Z_i|X_i)}p^Y_0(X_i)+o_p(1)\ .
	\end{align*}
	Therefore, 
	\begin{align*}
	&\sqrt{N}(\hat{\tau}-\tau)=\sqrt{N}\sum_{i=1}^N\left\{Z_i\frac{\hat{p}(X_i)}{\hat{\delta}^D(X_i)}Y_i-(1-Z_i)\frac{\hat{q}(X_i)}{\hat{\delta}^D(X_i)}Y_i-\tau\right\}\\
	=&\frac{1}{\sqrt{N}}\sum_{i=1}^N\bigg[\frac{2Z_i-1}{\delta^D(X_i)f_{Z|X}(Z_i|X_i)}Y_i-\tau\bigg]\\
	&-\frac{1}{\sqrt{N}}\sum_{i=1}^N\frac{p_1^Y(X_i)}{\delta^D(X)} \bigg\{\frac{Z_i}{f_{Z|X}(1|X_i)}-1\bigg\}\\
	&+\frac{1}{\sqrt{N}}\sum_{i=1}^N\frac{p_0^Y(X_i)}{\delta^D(X)} \bigg\{\frac{1-Z_i}{f_{Z|X}(0|X_i)}-1\bigg\}\\
	&-\frac{1}{\sqrt{N}}\sum_{i=1}^N\delta(X_i) \bigg\{\frac{2Z_i-1}{f_{Z|X}(Z|X_i)}\frac{D_i}{\delta^D(X_i)}-\frac{2Z_i-1}{f_{Z|X}(Z|X_i)}\frac{\mathbb{E}[D_i|Z_i,X_i]}{\delta^D(X_i)}\bigg\}+o_p(1) \quad \left[\text{since} \ \delta(X)=\frac{\delta^Y(X)}{\delta^D(X)}\right]\\
	=&\frac{1}{\sqrt{N}}\sum_{i=1}^N \varphi_{eff}(D_i,Z_i,X_i,Y_i)+o_p(1)
	\end{align*}
	where{\footnotesize
	\begin{align*}
	\varphi_{eff}(D_i,Z_i,X_i,Y_i)=&\frac{2Z_i-1}{f_{Z|X}(Z_i|X_i)}\frac{1}{\delta^D(X_i)}\bigg\{Y_i-D_i\delta(X_i)-\mathbb{E}[Y_i|Z_i=0,X_i] +\mathbb{E}[D_i|Z_i=0,X_i]\delta(X_i)\bigg\}+\delta(X_i)-\tau\ ,
	\end{align*}}
	is the efficient influence function given in \cite{wang2016bounded}.

%


\end{document}